# Sphaleron transition rate in the classical 1+1 dimensional abelian Higgs model at finite temperature

J. Smit and W.H. Tang [*]

Institute of Theoretical Physics, Valckenierstraat 65, 1018 XE Amsterdam, the Netherlands

We compute the sphaleron transition rate in the 1+1 dimensional abelian Higgs model at finite temperature, by real time simulation using the classical canonical ensemble.

## 1. INTRODUCTION

A nonperturbative calculation of the sphaleron transition rate in the Standard Model at finite temperature is difficult because it involves real time correlations. To deal with this the classical approximation was proposed and tested in the 1+1 dimensional abelian Higgs model [1], for which a semiclassical calculation of the rate was available [2]. The test involved the microcanonical ensemble and sphaleron transitions were counted 'by hand'. A nonperturbative definition of the rate

$$\Delta(t) \equiv \langle [C(t) - C(0)]^2 \rangle = \Gamma t, \quad \text{large } t, \qquad (1)$$

where $C(t)$ the Chern-Simons number, was used in a real time Langevin simulation [3]. This reproduced the temperature dependence and may be useful for a determination of the absolute rate [4]. Interesting results have also been obtained recently with a realistic heat bath method [5].

Here we shall present results using the canonical ensemble,

$$\begin{aligned}\Delta(t) &= Z^{-1} \int Dp\, Dq\, \exp[-H(p,q)/T] \\ &\quad [C(p(t),q(t)) - C(p,q)]^2. \end{aligned} \qquad (2)$$

The canonical variables $p(t)$ and $q(t)$ are evaluated with Hamilton's equations from the initial conditions $p(0) = p$, $q(0) = q$, with $p, q$ drawn from the ensemble.

The effective action of the abelian Higgs model at finite temperature will have approximately the classical form $S_{eff} \approx -\int dt\, dx\, [F_{\mu\nu}F^{\mu\nu}/4g^2 +$ $(\partial_\mu \phi - iA_\mu \phi)^*(\partial^\mu \phi - iA^\mu \phi) + m^2|\phi|^2 + \lambda|\phi|^4]$, at weak coupling $\lambda/m^2 \ll 1$, $g^2/\lambda = O(1)$, with temperature dependent parameters $m^2$, $\lambda$, $g^2$. This (quantum) effective action is used for the classical approximation.

In the Higgs phase $m^2 < 0$, $m_\phi^2 = -2m^2$, $m_A^2 = (g^2/2\lambda)m_\phi^2$, and the expectation value $\langle \phi \rangle \equiv v/2$, serves an expansion parameter $v^{-2} = 2\lambda/m_\phi^2$. The sphaleron energy is given by $E_s = 2v^2 m_\phi/3$.

At high temperatures $T/m_{A,\phi} \gg 1$, we may find classical behavior. Keeping the ratio $E_s/T$ fixed this means $v^{-2} \ll 1$, i.e. the semiclassical region.

To deal with the Rayleigh-Jeans divergence the classical field theory is put on a lattice, with lattice distance $a$ of order of the inverse temperature, $a = c/T$, with $c$ a numerical constant. It is convenient to make a scale transformation to dimensionless variables [3], $a \to a/v\sqrt{\lambda}$, $t \to t/v\sqrt{\lambda}$, $\phi \to v\phi$, $A_\mu \to v\sqrt{\lambda} A_\mu$, which gives an overall factor $v^2$ in the action. The scaled lattice action takes the form

$$\begin{aligned}S = \int dt\, a \sum_{n=0}^{N-1} &\{[(A_{t\,n+1} - A_{t\,n})/a - \partial_t A_{x\,n}]^2/2\xi \\ &+ |(\exp(-iaA_{x\,n})\phi_{n+1} - \phi_n)/a|^2 \\ &- |(\partial_t - iA_{t\,n})\phi_n|^2 + (|\phi_n|^2 - 1/2)^2\}, \quad(3)\end{aligned}$$

where $\xi = g^2/\lambda$ and we assume periodic boundary conditions with spatial size $L = Na$. The hamiltonian and sphaleron energy scale as $H \to v^3\sqrt{\lambda}H$, $E_s \to v^3\sqrt{\lambda}E_s$. It is convenient to scale also the temperature $T \to v^3\sqrt{\lambda}T$, such that the Boltzmann factor keeps its usual form. In scaled variables $\langle\phi\rangle = 1/\sqrt{2}$, $E_s = 2\sqrt{2}/3 \approx 0.94$,

---
[*] Speaker at the conference

$m_\phi = \sqrt{2}$, $m_A = \sqrt{\xi}$.

A crucial observation is that the scaled lattice distance $a$ goes to zero in the semiclassical limit: under the scaling we have $c = aT \to c = aTv^2$. Hence $a \propto v^{-2}$ at fixed scaled temperature (fixed $E_s/T$). In the classical approximation $v^2$ drops out, but its role is taken over by the scaled lattice distance.

Following ref. [3] we use the Coulomb gauge $\partial_x A_x = 0$ and solve for the Gauss constraint. Then the Chern-Simons number

$$C = \int dx\, A_x/2\pi = LA_x/2\pi \qquad (4)$$

can be taken as one of the canonical variables. The hamiltonian follows as usual.

## 2. NUMERICAL STRATEGY

We use a second order Langevin procedure to generate the canonical ensemble of $p$'s and $q$'s. Following ref. [3] we use polar coordinates $\phi = \rho \exp(i\theta)$ and apply the random Langevin forces only to the gauge invariant variables $\rho$, $p_\rho$, and not to the gauge variant $\theta$, $p_\theta$. The random forces are also not applied to $C$ and $p_C$, which makes it possible to monitor thermalization of $p_C$. The system is kicked at times $t = kh$ (integer $k$), and for $H = \sum_n p_{\rho n}^2/4a + \cdots$ the algorithm can be presented as

$$\rho_n(t+h) = \rho_n(t) + \int_t^{t+h} dt' \frac{\partial H}{\partial p_{\rho n}}, \qquad (5)$$

$$p_{\rho n}(t+h) = p_{\rho n}(t) + \sqrt{4afhT}\,\eta_n(t) - \int_t^{t+h} dt' [\frac{\partial H}{\partial \rho_n} + fp_{\rho n}], \qquad (6)$$

where $f$ is the friction coefficient, $T$ is the specified ('input') temperature and the $\eta$ are independent gaussian random numbers with unit variance and zero mean. The $\eta$ are absent from the other dynamical equations. The singularity of polar coordinates at $\rho = 0$ led us to use a variable stepsize leapfrog algorithm for the integration of Hamilton's equations over the interval $(t, t+h)$.

The finite Langevin stepsize $h$ introduces an error which we monitored by requiring the 'output temperatures' $T_C$ and $T_\rho$,

$$\langle \frac{\xi L p_C^2}{4\pi^2} \rangle = T_C, \quad \langle \frac{p_\rho^2}{2a} \rangle = T_\rho, \qquad (7)$$

to be such that their inverse $\beta_{C,\rho} = 1/T_{C,\rho}$ differed less than 0.1 in absolute value from $\beta = 1/T$. This criterion is based on the interpretation that Langevin errors lead to an effective temperature, the output temperature. Then the expectation $\Gamma \propto \exp(-\beta E_s)$, suggests that an absolute error in $\beta$ of $\approx 0.1$, causes a relative error in $\Gamma$ of $\approx 10\%$.

Having produced an independent configuration of $p$'s and $q$'s we took this as the initial condition for the real time integration of Hamilton's equations. For this we could use the original cartesian coordinates, as the condition of charge zero is easily satisfied by projecting regularly onto zero charge (this only involves changes of machine precision order). After real time integration the Langevin process was started again and the process was repeated untill sufficient statistics was obtained.

Our configurations were actually more microcanonical than canonical, because the Langevin processes were stopped (for historical reasons) at times that the total energy had its mean value. As expected we found in a few checks that the true canonical ensemble gives the same results within errors. The friction parameter was taken $f = 1$, and we checked that results do not depend on $f$, as expected.

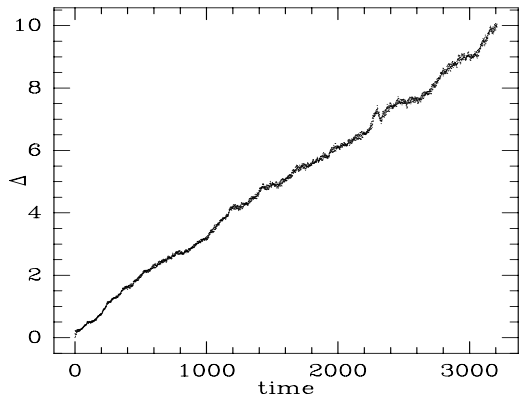

Figure 1. Diffusion $\Delta(t)$ for $\beta = 11$, $\xi = 4$, $L = 16$, $a = 0.32$.

For low temperatures $\Delta$ tends to be small and large real times $t$ were needed. We required $\Delta > 10$. An accurate fifth order Adams-Bashforth-Moulton predictor-corrector multi-step integration algorithm was used to keep the numerical drift in the total energy sufficiently small. Otherwise the diffusion $\Delta$ will not be linear in $t$ (cf. fig. 1).

## 3. RESULTS

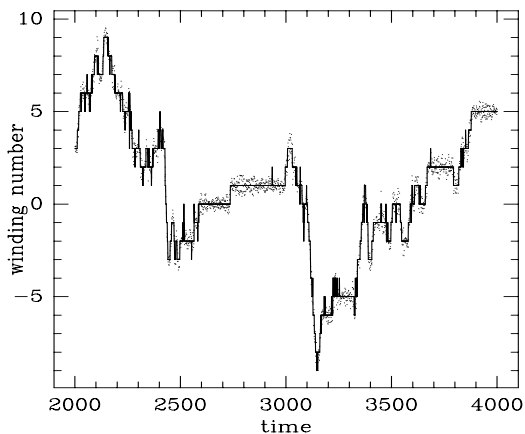

Figure 2. Winding number (line) and Chern-Simons number (dots) for $\beta = 8$, $\xi = 4$, $L = 16$, and $a = 0.32$, during Hamilton evolution.

Fig. 2 shows an example of the real time behavior of $C$ at low temperature. A comparison is made with the winding number of the scalar field $w = -\sum_n \partial \theta_n$, with $\partial \theta_n \equiv \theta_{n+1} - \theta_n$ (mod) $2\pi \in (-\pi, \pi]$. On the lattice the winding and Chern-Simons numbers are defined modulo $N$. Fig. 1 shows an example of $\Delta(t)$ at low temperature. In fig. 3 a comparison is made with the analytic semiclassical result

$$\frac{\Gamma}{m_\phi^2 L} = \frac{(3)^{1/2}}{(2\pi)^{3/2}} (\beta E_s)^{1/2} \exp(-\beta E_s) \quad (8)$$

$$\left[ (s+1) \frac{\Gamma(\alpha+s+1)\Gamma(\alpha-s)}{\Gamma(\alpha+1)\Gamma(\alpha)} \right]^{1/2},$$

where $\alpha = \sqrt{2\xi}$ and $s = (-1 + \sqrt{1+8\xi})/2$. The value of $\xi$ is again $\xi = 4$. The upper data are for $a = 0.32$, the lower data for $a = 0.16$ (at $\beta = 10, 11$ for $a = 0.32$ only). The system size is $L = 16$ ($N = 50$ and $N = 100$), and we obtained the same results within errors for $L = 10.28$, for a number of $\beta$ values. However, for sizes as low as $L = 8$ we see clear deviations.

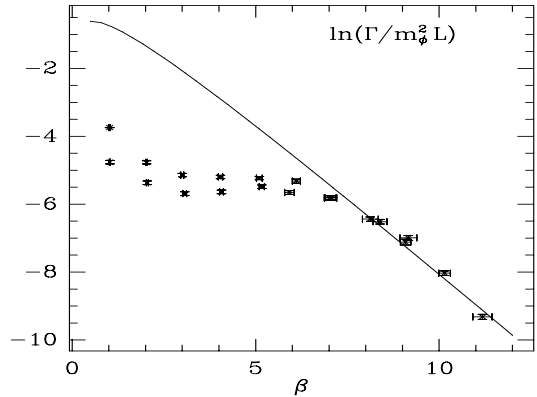

Figure 3. Results for $\ln(\Gamma/m_\phi^2 L)$. The solid line represents the analytic formula (8).

The errors in the rate $\Gamma$ are statistical and determined with the jackknife method from at least 900 configurations. For $\beta$ we used the output temperature (7), with errors obtained by the binning method. The input $\beta$ is the nearest integer.

We conclude from fig. 3 that the classical simulation is able to reproduce the semi-classical formula for $\beta \gtrsim 7$. The data at smaller $\beta$ depend on the lattice distance $a$ and appear therefore to be ambiguous. Perhaps smaller $a$ are needed in this region of $\beta$ values; or the classical or the semi-classical approximation or both break down.

**Acknowledgements:** We would like to thank A.I. Bochkarev for useful discussions. This work is financially supported by the Stichting voor Fundamenteel Onderzoek der Materie (FOM).